\documentclass[preprint,showpacs,superscriptaddress,floatfix,pre]{revtex4}

\usepackage{amsmath}
\usepackage[dvips]{graphicx}
\usepackage{epsfig}
\usepackage{tabularx}

\begin{document}

\title{A kg-mass prototype demonstrator for DUAL gravitational
wave detector: opto-mechanical excitation and cooling.}

\author{M. Anderlini, F. Marino and F. Marin}

\affiliation{Dipartimento di Fisica, Universit\`a di Firenze,
INFN, Sezione di Firenze, and LENS \\ Via Sansone 1, I-50019 Sesto
Fiorentino (FI), Italy}

\email[Corresponding author: ]{marin@fi.infn.it}

\date{\today}
\begin{abstract}

The next generation of gravitational wave (gw) detectors is
expected to fully enter into the quantum regime of force and
displacement detection. With this aim, it is important to scale up
the experiments on opto-mechanical effects from the microscopic
regime to large mass systems and test the schemes that should be
applied to reach the quantum regime of detection. In this work we
present the experimental characterization of a prototype of
massive gw detector, composed of two oscillators with a mass of
the order of the kg, whose distance is read by a high finesse
optical cavity. The mechanical response function is measured by
exciting the oscillators though modulated radiation pressure. We
demonstrate two effects crucial for the next generation of
massive, cryogenic gw detectors (DUAL detectors): a) the reduction
of the contribution of 'local' susceptibility thanks to an average
over a large interrogation area. Such effect is measured on the
photo-thermal response thanks to the first implementation of a
folded-Fabry-Perot cavity; b) the 'back-action reduction' due to
negative interference between acoustic modes. Moreover, we
obtain the active cooling of an
oscillation mode through radiation pressure, on the described
mechanical device which is several orders of magnitude heavier
than previously demonstrated
radiation-pressure cooled systems.

\end{abstract}


\maketitle

\section{Introduction}

Ground-based gravitational wave (gw) detectors are
traditionally classified into large baseline interferometers and
resonant antennas. Recently, a new class of gw detectors has been
proposed, based on huge masses kept at cryogenic temperature and
called DUAL detectors~\cite{Cerdonio}-\cite{Marin}. At difference
from previous massive cryogenic antennas (such as Weber bars), the
DUAL system do not aim at reaching the best peak sensitivity in a
narrow band around a resonance frequency, but it takes advantage
of elastic forces to achieve a useful sensitivity in a wide
frequency range. For this purpose, it is no longer equipped with
resonant mechanical amplifiers, and it needs a very sensitive
readout. The goal is realizing a detector which covers the
acoustic frequency region (1-5~kHz).

A DUAL gw detector exploits two oscillation modes of a mechanical
system with a readout symmetric with respect to the center of
mass. Do to the geometry, the responses to the readout force of
the two modes must be summed, while the responses to the tidal
force of the gw are subtracted. In the frequency region between
the two resonance frequencies, the susceptibilities of the two
modes are in anti-phase, giving a reduced response to the readout
force and an enhanced response to the gw~\cite{Briant,Bonaldi}
('back-action reduction'). Such behavior has been recently
verified and explored experimentally in Ref.~\cite{Caniard}, where
the authors study the
mechanical response to radiation pressure of a cm-size
plano-convex mirror which is part of a high Finesse optical
cavity. The internal modes of the mirror analyzed are in the $\sim
10^6$~Hz frequency and $\sim 10^{-3}$~kg mass range. It is worth pointing out that the
'back-action reduction' mechanism is also the basis for a
proposed scheme of quantum non-demolition measurement.

Two further important effects are considered in the design of
DUAL: a) a clever readout geometry allows to be blind to some
low-frequency acoustic modes that would bring thermal and
back-action noise peaks in the useful frequency range (`mode
selectivity'); b) a broad interrogation area enables to reduce the effect of 'local'
deformations due to thermal motion and readout force, i.e., from a
different point of view, the effect of high frequency modes ('wide
area readout').

Different configurations of DUAL have been proposed and studied:
two nested spheres~\cite{Cerdonio}, where the relevant modes are
the first quadrupolar mode of the inner and outer sphere; two
nested cylinders~\cite{Bonaldi}, again acting on the first
quadrupolar mode of the nested bodies; a single hollow
cylinder~\cite{Bonaldi1}, exploiting its first and second
quadrupolar modes; a symmetric set of cylinders~\cite{Marin},
where the first DUAL mode is given by the elastic link
between them and the second one by their first oscillation mode.

For any configuration, the possibility of an optical readout is
extremely useful for its potentiality to reach a detection
sensitivity at the standard quantum limit\cite{Caves}, or even
surpass it\cite{Belfi}. An optical scheme expressly conceived for
the application in a DUAL detector is proposed and studied in
Ref.~\cite{Marin1}. The idea is 'folding' a Fabry-Perot cavity in
order to interrogate a large detector area with a beam bouncing
several times between the sensing masses, before being reflected
back on itself by the end mirror to close the cavity. Such a kind
of folded Fabry-Perot (FFP) has never been implemented
experimentally.

An interesting possibility for boosting the sensitivity of massive
detectors to short bursts is suggested by Vitali {\it et
al.}\cite{Vitali02}. They show that, in some conditions, periodic
cycles of fast feedback cooling of the sensitive mechanical mode,
followed by a measurement period lasting a fraction of the
temperature recovery time, brings an improvement of the
sensitivity by even an order of magnitude.

Cooling of a mechanical oscillator by means of radiation
pressure\cite{Mancini} has been recently demonstrated on
micro-resonators using both active techniques ('cold
damping')\cite{Cohadon99,Kleckner06}, with feedback acting on the
optical power impinging on the mirror, or passive schemes ('self
cooling'), exploiting the mechanical effect of red-detuned laser
radiation~\cite{Gigan06,Arcizet06,Schliesser06,Thompson08}. The
above results were obtained on effective oscillator masses of less
than 1~g, and frequencies in the $10^4-10^8$~Hz range. A recent
work~\cite{Mow08} describes the cooling of a gram-size mirror held by
a cantilever flexure, with a resonance frequency of $\sim$100~Hz.
Here the cantilever damping is obtained through a so-called hybrid
scheme, exploiting active feedback on the cavity length which
actually modifies the radiation pressure exerted by detuned laser
radiation. The cooling effect can be attributed to the optical
spring originated by radiation pressure, modified (with an added
reactive component) by feedback. A similar hybrid technique is used also by Corbitt {\it et al.}\cite{Corbitt} to cool a suspended $\sim$1~g mirror, with the difference that here the detuning is modified by feedback acting on the laser frequency instead of the cavity length. However, the system explored in Ref.~\cite{Corbitt} is very peculiar: the optical spring is so strong that it completely determine the oscillation frequency (around $\sim$1~kHz) and the mirror confinement is provided by the optical potential rather than elastic forces.

A completely different cooling technique has been recently applied
to the $\sim1$~kHz fundamental mode of the 3~tons aluminum bar of
the AURIGA detector\cite{Zendri08}. In that work the feedback is
electrical  and is applied to the capacitor of the mechanical
readout. Starting from a cryostat temperature of 4.2~K, the authors obtain a record temperature of 0.17~mK.

Due to the innovative DUAL concept, on the one hand a preliminary experimental
study on a small-scale prototype is required. On the other hand,
it is important to scale up the experiments on opto-mechanical
effects from the microscopic to the macroscopic range of masses
and in the kHz frequencies, and test the schemes that should be
applied to reach the quantum regime of detection. In this work we
present the experimental characterization of a prototype of DUAL
detector, composed of two oscillators with a mass of the order of
the kg, whose distance is read by a high finesse optical cavity.
The mechanical response function is measured by exciting the
oscillators through modulated radiation pressure. We demonstrate
two of the above described effects crucial for the DUAL detectors:
a) the reduction of the contribution of 'local' susceptibility due
to average over a large interrogation area. Such effect is
measured on the photo-thermal response\cite{DeRosa02,DeRosa06}
thanks to the first implementation of a folded-Fabry-Perot
cavity\cite{Marin1}; b) the 'back-action reduction' due to
negative interference between acoustic modes. Moreover, we obtain
the active cooling of an oscillation mode through radiation
pressure, on our device, which is orders of
magnitude heavier than any previously demonstrated
radiation-pressure cooled system.

\section{Experimental setup}
\label{Expsetup}

\begin{figure}[t]
\begin{center}
\includegraphics*[width=0.9\columnwidth]{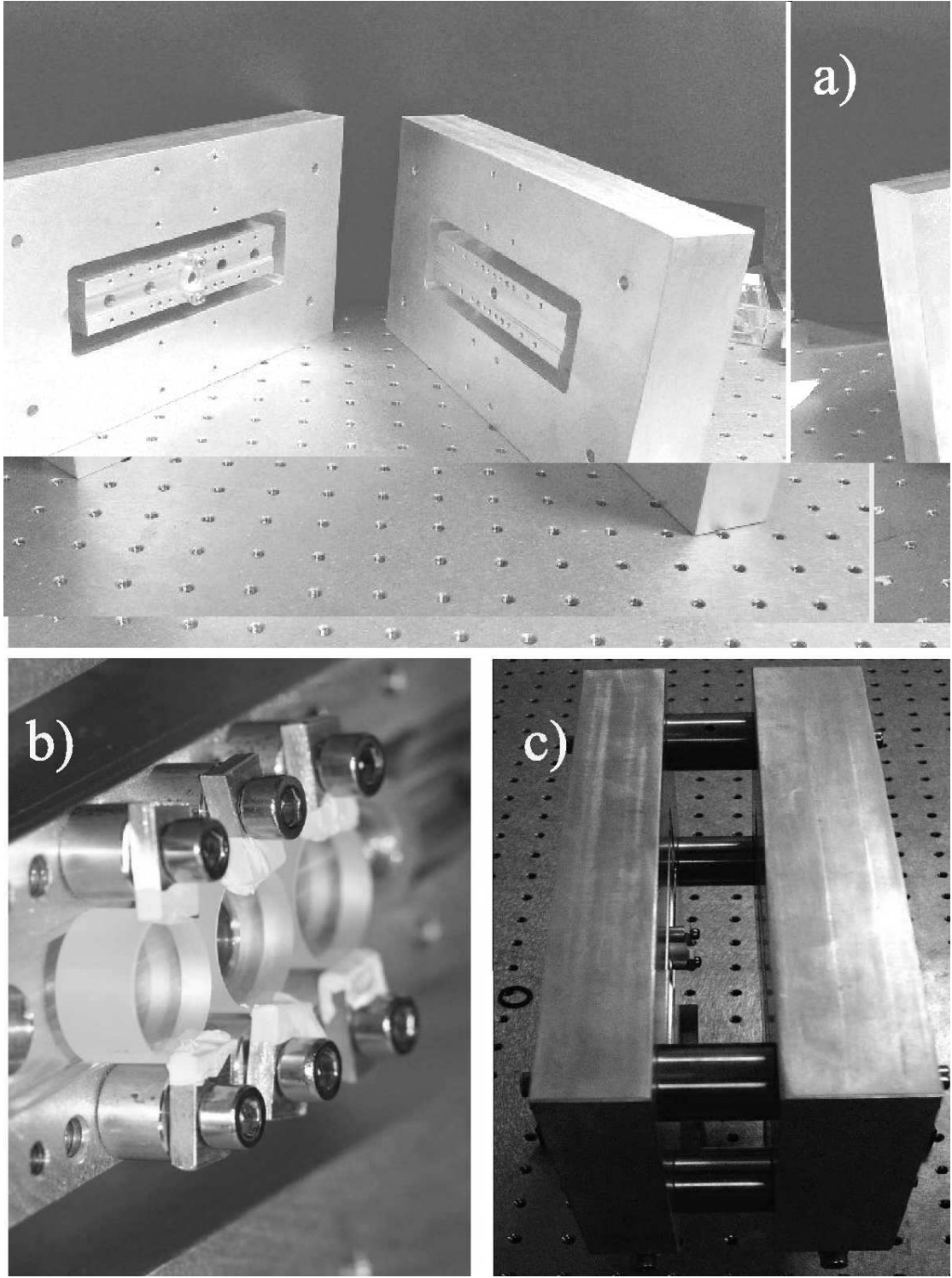}
\end{center}
\caption{Photos of our prototype double oscillator with folded
Fabry-Perot readout. a) The two frames, with the oscillating
masses in the center; the masses are connected to the frames by
leaving lateral membranes on the back of the frames. b) Three
mirrors of a row of five are fixed on one mass (in Figs. a) and c)
one mirror is left as reference). c) The device is assembled with
INVAR spacers.} 
\label{foto}
\end{figure}

The oscillators are made with two 135~mm wide, 30~mm high aluminum masses (test masses), fixed to frames along their short edge by $\sim$mm
thick lateral membranes. The first mass
is 30~mm thick and 0.33~kg heavy; the
second one is 40~mm thick and 0.44~kg heavy. The
two oscillators are carved from single aluminum blocks, whose external parts are kept
together by INVAR spacers and represent
the external frames of the structure (see Fig.~(\ref{foto})). Two rows of
12.7~mm diameter mirrors form the FFP:  five on one mass, including a flat,
130~ppm transmission  input mirror and a 1~m radius end mirror
angled by 18$^{\circ}$, and four flat mirrors on the opposite
mass. The distance between opposite mirror surfaces
is $D=20$~mm. The input mirror is positioned on the outer side of the 30~mm
thick mass with respect to the other mirrors of the row, for
easier alignment. The FFP cavity length is $L=200$~mm. Due to the
non-normal reflections, for each longitudinal mode the
s-polarization resonance frequency is detuned by 6~MHz from the
p-polarization resonance (the mirror coating is optimized for
normal incidence), with no appreciable linewidth difference. The
prototype is placed on a cantilever mechanical suspension in a
thermally stabilized vacuum chamber.

The mechanical properties of the two coupled oscillators have been
characterized by positioning accelerometers on the two frames and
analyzing the response to a global mechanical excitation. A
comparison with a Finite Element Method (FEM) model allows us to
identify the peaks corresponding to the translation and torsion
modes of the two masses (their shapes are shown in Fig.~(\ref{spettro}). In particular, the translation modes are
found at $\sim$800~Hz (for the heavier mass) and $\sim$1200~Hz
(for the lighter mass). The same FEM analysis gives the effective
masses of the translation modes, excited and read on a small circular area corresponding to the central mirrors. The results are respectively 0.31~kg and 0.41~kg, not far from the physical masses.

\begin{figure}[t]
\begin{center}
\includegraphics*[width=0.9\columnwidth]{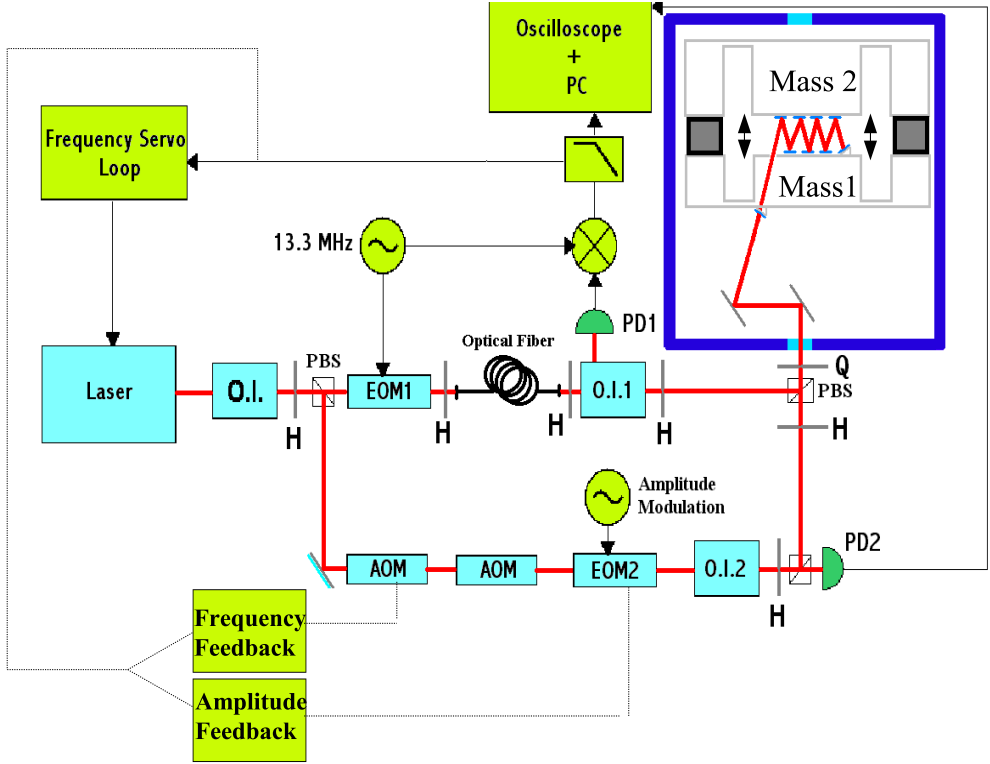}
\end{center}
\caption{Scheme of the experimental apparatus. O.I.: optical
isolator; AOM: acousto-optic modulator; EOM: electro-optic
modulator; H: half-wave plate; Q: quarter-wave plate; PD:
photodiode; PBS: polarizing beam-splitter. The double arrows indicate the translation motion of the test masses. The servo loops shown with dashed lines are only used in the cooling experiment, alternatively in the two configurations (with feedback acting in the frequency or in the amplitude of the cooling beam).} 
\label{apparato}
\end{figure}

The experimental set-up for the opto-mechanical characterization
of the prototype is sketched in Fig.~(\ref{apparato}). The light
source is a cw Nd:YAG laser working at 1064~nm. After a 40~dB
optical isolator, the laser radiation is split into two beams. On
the first one, a resonant electro-optic modulator (EOM1) provides
phase modulation at 13.3~MHz with a depth of about 1~rad used for
the Pound-Drever-Hall\cite{Drever} (PDH) detection scheme. The
light is then transmitted by a polarization maintaining optical
fiber and a second optical isolator (O.I.1). The second beam can be
shifted in frequency by means of two acousto-optic modulators
(AOM) and modulated in amplitude by an EOM (EOM2) followed by an
optical isolator (O.I.2). We use respectively the +1 and -1
diffracted orders of the AOMs, so that the total frequency
displacement corresponds to the difference of the AOM frequencies
and can be tuned by several MHz around zero. After O.I.2, part
of the second beam is detected by a photodiode (PD2) for
monitoring its amplitude modulation. The two beams are combined
with orthogonal polarizations in a polarizing beam-splitter and
sent to the folded optical cavity. A quarter-wave plate allows us to
optimize the matching of the polarization to the cavity modes.
The reflected first beam, on his back path, is deviated by the
input polarizer of the O.I.1 and collected by a photodiode (PD1)
for the PDH detection. This PDH signal is used for laser frequency
locking, while the second beam is tuned across the resonance using
the AOMs. Even when both beams are resonant with respect to the
respective polarization modes, their frequency splitting allows to
eliminate any spurious interference and cross-talk.

The displacement noise spectrum has been obtained directly from
the Pound-Drever signal with the laser weakly locked on the
cavity, and stopping the second beam. The signal is calibrated by
modulating the laser frequency through its internal piezoelectric
crystal, with a depth and a modulation frequency smaller than the
FFP cavity linewidth, and using a phase-sensitive detection on the
PDH signal. The laser piezoelectric crystal is itself calibrated
by observing the sidebands at 13.3~MHz on a wide, slow enough
scan. The PDH calibration procedure, performed varying the
modulation frequency, allows also to correct the data for the
servo loop.

In order to characterize the response of the FFP to variations of
the intracavity power, the laser is weakly locked on the cavity
using the first beam while the second beam is amplitude modulated
at different frequencies. The extraction of the cavity response is
obtained again from the PDH signal. The signal from PD2 and PD1
are acquired simultaneously with a digital oscilloscope, and
successively elaborated to infer the component synchronous with
the modulation. The depth of the intracavity power modulation is
calculated from the signal of PD2 by taking into account the mode
matching and the cavity coupling factor. The latter is inferred
from the depth of the dip in the reflected beam when scanning on
the resonance, and it is consistent with the independent
measurements of cavity Finesse and input mirror transmission.

A first important test concerns the optical properties of the FFP
configuration, that had not been implemented before. We could
indeed obtain a mode-matching of 93\% (comparable with that
obtained with a simple cavity) and a Finesse of 6000 (linewidth
125~kHz) with an intracavity optical power of the order of 1~W, in
agreement with a calculation based on the independently measured
losses of the mirrors.

\section{Opto-mechanical characterization}

\subsection{Displacement noise spectrum}

\begin{figure}[t]
\begin{center}
\includegraphics*[width=0.9\columnwidth]{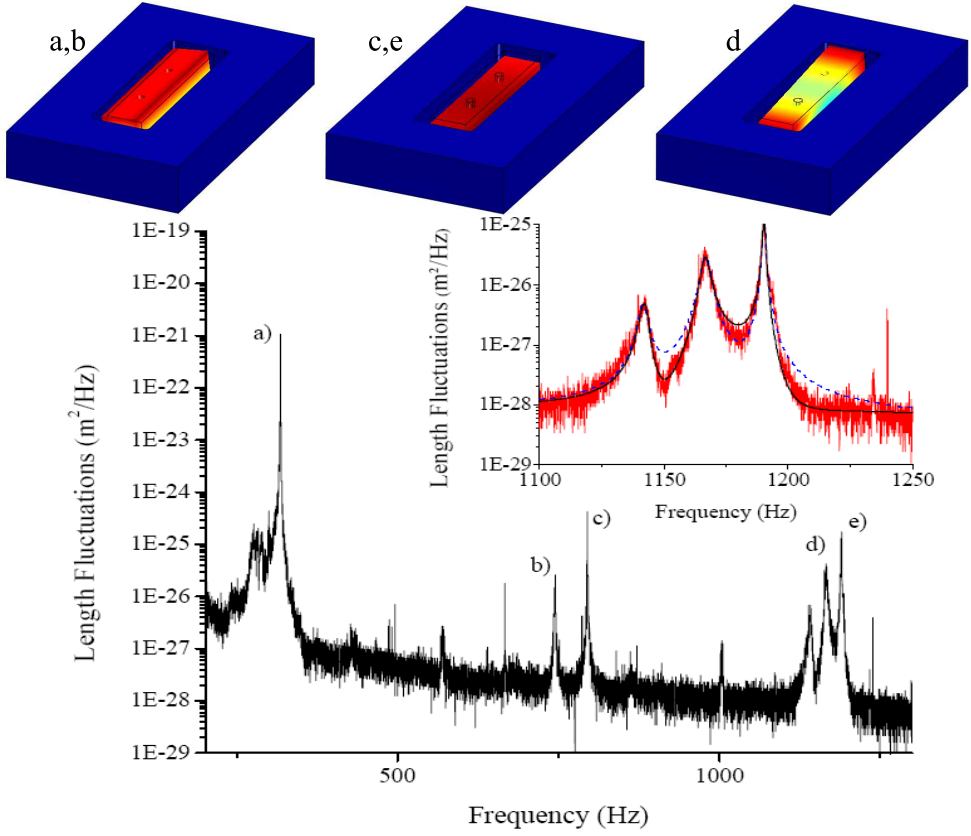}
\end{center}
\caption{Spectral density of the cavity length fluctuations. Some
of the peaks are attributed to particular modes by comparison with
the FEM model, namely translation modes (c and e) and torsion modes
(a, b and d). Their structure is shown in the upper part of figure. For the translation modes, it is meaningful to
attribute the signal to fluctuations in the distance $D$ between
the two masses, according to Eq.~(\ref{DtoL}). In this way, the
spectral density shown in the graph is reduced by a factor of $64
\cos^2\theta\simeq 58$. In the inset, fit with thermal noise as in
Eq.~(\ref{noise1}) (dashed line) and external noise as in
Eq.~(\ref{noise2}) (solid line).} 
\label{spettro}
\end{figure}

The cavity length noise spectrum evidences the mechanical modes
directly coupled to the optical cavity (i.e., modes changing the
intracavity optical path). Among them, it is possible to recognize
the lower order translation and torsion modes, previously
identified in the mechanical characterization of the cavity
(Fig.~(\ref{spettro})). These mechanical resonances emerge from a
background given by the free-running laser frequency noise (in the
frequency region of interest, it is approximately
$10^8/f^2$~Hz$^2$/Hz where $f$ is the frequency expressed in Hz).

\begin{figure}[t]
\begin{center}
\includegraphics*[width=0.9\columnwidth]{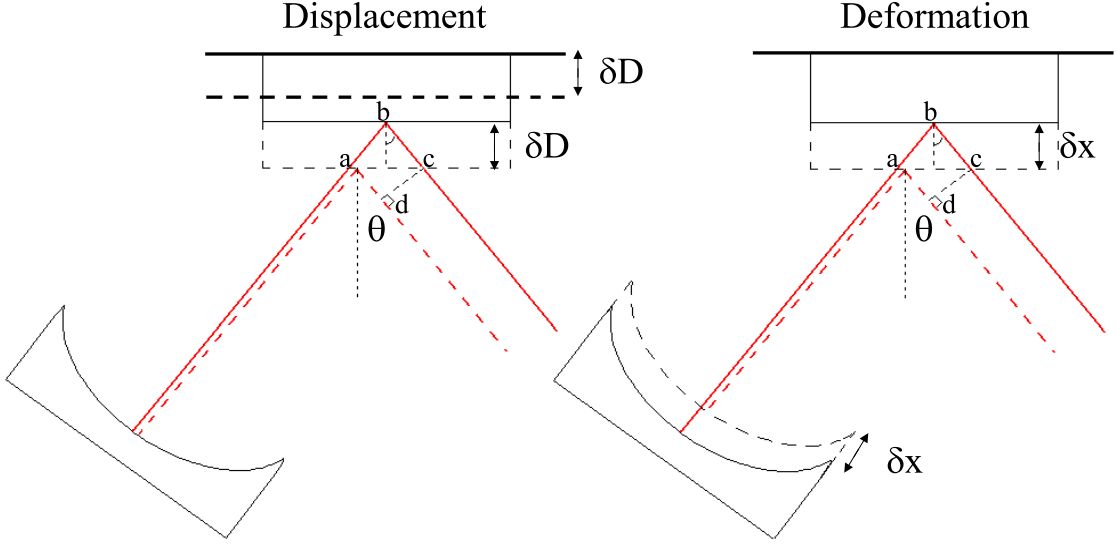}
\end{center}
\caption{Scheme of the optical length change in the case of a mass displacement $\delta D$ (left) or a photo-thermal deformation $\delta x$ (right). In the case of the folding mirror, the path change is (considering the right panel) $(ad - abc)=2\cos\theta \, \delta x$, while for the end mirror it is simply $\delta x$. The incidence angle ($18^{\circ}$ in our setup) is exaggerated for clearness.} 
\label{spostamenti}
\end{figure}

The vertical scale in Fig.~(\ref{spettro}) is expressed in terms
of spectral density of variations $\delta L$ of the cavity optical
length $L$ (measured from the input mirror to the end mirror),
which are directly inferred from the measured spectrum using the
calibration procedure described in the previous section. For an intuitive understanding, the system can be approximated by infinitely rigid masses linked with mass-less membranes to a still frame. With this scheme, the translation modes corresponds to pure translations of the masses on the cavity plane and
perpendicularly to the mirror surfaces, as shown by arrows in Fig.~(\ref{apparato}). In this case, simple geometrical
considerations (illustrated in Fig.~(\ref{spostamenti})) lead to the relation
\begin{equation}
\label{DtoL} 
\delta L = 8\cos\theta \, \delta D
\end{equation}
where $\delta D$ is the (small) variation of the distance between
the two masses, $\delta L$ is the corresponding change in the
cavity optical length, and $\theta$ is the incidence angle on the
folding mirrors ($18^{\circ}$ in our case). The expression
(\ref{DtoL}), inverted, can be used to convert the measured
displacement from $\delta L$ to $\delta D$. In this way, from the
measurement in Fig.~(\ref{spettro}) we can directly deduce the
displacement of 'ideal' translation modes, while for other
mechanical modes one can only see in the spectrum the
corresponding fluctuations of the cavity length. In the approximated scheme with rigid masses, the relevant torsion modes correspond to rotations of the mass around an horizontal (modes a,b in Fig.~(\ref{spettro})) or a vertical axis (mode d in the figure). In this approximation, such modes do not change the cavity length at the first order in the angular motion. Therefore, the system should be completely blind to the torsion modes, that are excited and detected only thanks to deformations in the masses and frames.  

The mechanical quality factors of the modes, measured from their
spectral width, are respectively 1800 (heavy mass translation mode
at 795~Hz), 850 (light mass translation mode at 1190~Hz), and 310
(heavy mass torsion mode around a vertical axis, at 1166~Hz).

The ratio between the areas of the two peaks corresponding to the
first order translation mode of the two masses is 1.45, in fair
agreement with the value of 1.69 calculated by assuming a white driving force noise. A modal effective
temperature $T_{eff}$ can be calculated from these areas,
according to
\begin{equation}
\label{Teff} \int S_{x}\, \textrm{d}\nu =
\frac{kT_{eff}}{M\omega_0^2}
\end{equation}
where $S_x$ is the displacement noise spectral density, $k$ is the
Boltzmann constant, $M$ is the modal mass and $\omega_0/2\pi$ its
eigenfrequency. Using the values for the mass calculated by the
FEM analysis, we obtain for the two modes effective temperatures
of about 3300~K and 3900~K respectively, clearly showing that the
effects of external, technical noise is about one order of
magnitude larger than thermal noise. We can infer from this
measurement an horizontal displacement noise, at the level of the
FFP frames, of about $2\cdot 10^{-33}$~m$^2$/Hz. This is
compatible with seismic noise around 1~kHz, filtered by the optical
table and the FFP suspension.

The excess external noise is also confirmed by the spectral shape
around 1.2~kHz, where we find three nearly-degenerate modes: as
already mentioned the highest frequency ($\sim$1190~Hz) belongs to
the translation mode of the light mass, the intermediate frequency
($\sim$1166~Hz) is a torsion mode of the heavy mass,
while the lowest frequency mode (around $\sim$1142~Hz) remains
unidentified. In the case of thermal noise excitation, the
contributions of all these modes should add incoherently in the
noise spectrum, that should read\cite{Saulson,nota}:
\begin{equation}
\label{noise1} 
S_x = \frac{4kT}{\omega}\sum_i \textrm{Im} \chi_i=
4kT \sum_i \frac{\omega_i}{Q_i
M_i\left[(\omega_i^2-\omega^2)^2+\left(\frac{\omega\omega_i}{Q_i}\right)^2\right]}
\end{equation}
where for each mode $\chi_i$ is the susceptibility, $M_i$ the
mass, $\omega_i$ the eigenfrequency and $Q_i$ the quality factor,
and $T$ is the thermodynamic temperature. On the other hand, an
external noise $S_x^{ext}$ should excite them coherently, and
therefore be filtered by a superposition of modes response
functions, such as:
\begin{equation}
\label{noise2} 
S_x = \left| \sum_i \chi_i\right|^2 S_x^{ext}=
\left| \sum_i
\frac{1}{M_i\left[\omega_i^2-\omega^2-i\frac{\omega\omega_i}{Q_i}\right]}\right|^2
S_x^{ext}
\end{equation}
where the effective masses do not necessarily correspond to the
previous ones. 

We have fitted the experimental spectrum, in the mentioned region around 1.2~kHz, with the expressions (\ref{noise1}) and (\ref{noise2}). We remark that  the two fitting procedures uses the same number of free parameters, namely the set $\{Q_i, \omega_i\}_{i=1,3}$, three amplitudes, and the coefficient of the frequency noise background. Concerning the amplitude, we have chosen to fix the mass of the translation mode at its FEM value, leaving the other two masses as free and multiplying by an overall factor that can be identified with $S_x^{ext}$ of expression (\ref{noise2}) or with the temperature in expression (\ref{noise1}). The two fitting curves are reported in the inset of Fig.~(\ref{spettro}). The fit with
expression (\ref{noise2}) is clearly better. A quantitative evaluation, underlining the contribution of the wings of the peaks where the difference between the two fitting functions is more evident, is obtained by calculating the quadratic sum of differences between the logarithm of the experimental data and of the corresponding theoretical values (in other words, the $\chi^2$ in logarithmic scale). This indicator is a factor of 1.7 higher when using expression (\ref{noise1}), with respect to Eq.~(\ref{noise2}). Since the number of free parameters is the same, the analysis gives a strong indication in favor of the model behind Eq.~(\ref{noise2}), with a flat external noise spectrum. Of course, this is confirmed by the unrealistically high temperature (or low mass) that would be required by expression (\ref{noise1}). However, we remark here that, in the case of multiple overlapping peaks, there is not merely an increased spectral power due to extra-noise, but also a change in the shape of the spectrum that can be a stronger, calibration-independent indication.

\subsection{Response to modulation of the intracavity power}

\begin{figure}[t]
\begin{center}
\includegraphics*[width=0.9\columnwidth]{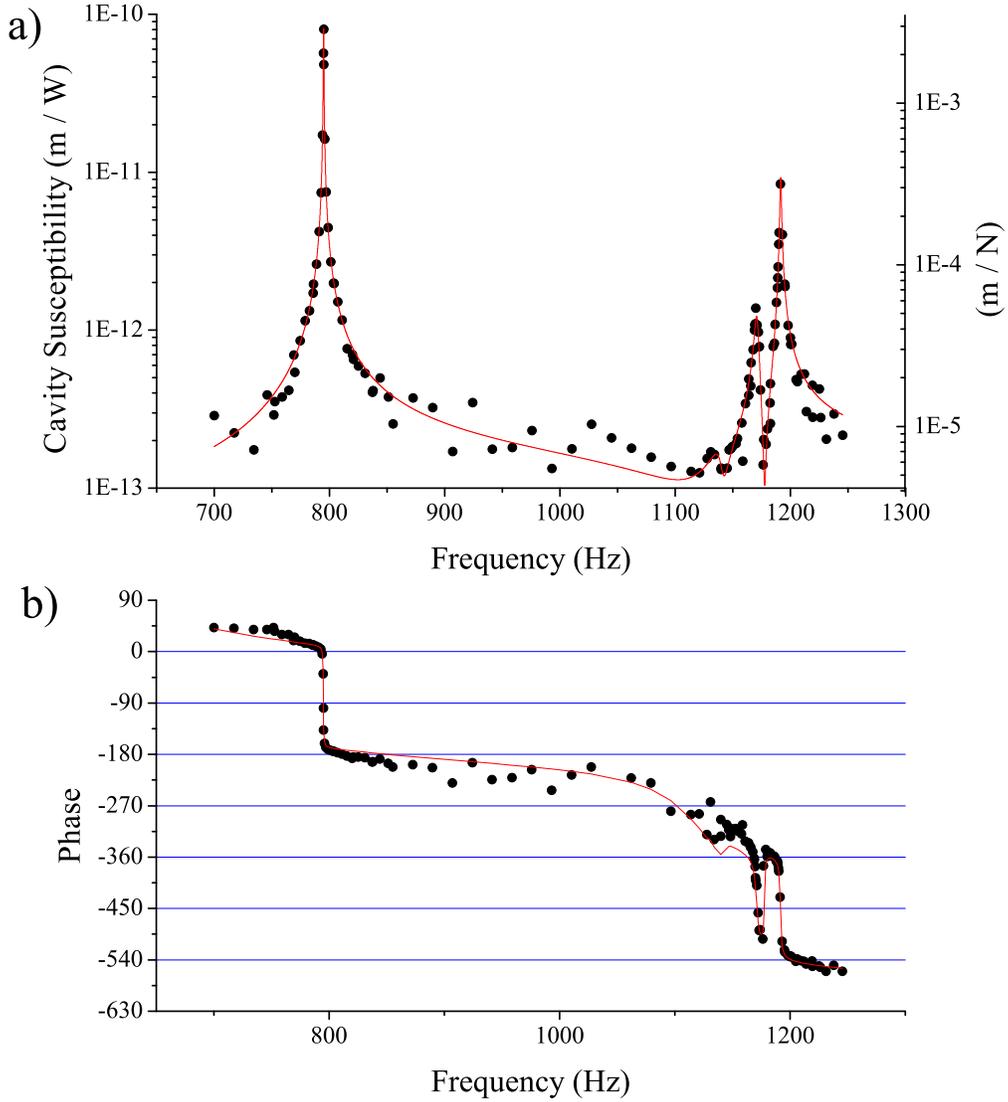}
\end{center}
\caption{Amplitude (upper graph) and phase (lower graph) of the
response of the cavity length to modulation of intracavity power. The solid line is the result of a fitting procedure based on the expression (\ref{fitfun}).}
\label{risposta}
\end{figure}

In Fig.~(\ref{risposta}) we report the phase and amplitude
response of the FFP to a modulation of the intracavity radiation
intensity. The measured signal is reported in terms of changes in
the cavity length $L$. To obtain the response, the signal is
divided by the amplitude of the modulation in the intracavity
radiation pressure, giving a susceptibility in m/N. 

Following the discussion in the previous section, the derived quantity has an intuitive interpretation only for the 'ideal' translation modes of the test masses. In this case, the relation (\ref{DtoL}) allows to find the variations
of the mass positions and, as we will see later, to extract values of the modal masses directly comparable with the FEM model and/or the physical masses.

The mechanical resonances emerge on top of a (nearly) constant
background associated with photo-thermal effect, i.e., to mirrors thermal
expansion due to heating by the absorbed intracavity
radiation~\cite{Brag1,Cerdonio1}. This effect is more meaningfully
expressed in terms of m/W, as indicated on the left vertical axis
of  Fig.~(\ref{risposta}). As will be discussed later in this section the quantitative explanation of the
photo-thermal background implies the verification of the 'wide
area readout' effect.

The complete response, between 700~Hz and 1250~Hz, is fitted to
the coherent superposition of a constant, complex background $C_{phth}$ and four mechanical
resonance, using the following function:
\begin{equation}
\label{fitfun}
X(\omega) = \sum_{i=1,4}
\frac{A}{M_i\left[\omega_i^2-\omega^2-i\frac{\omega\omega_i}{Q_i}\right]}+C_{phth} \, .
\end{equation}
The fitting procedure considers at the same time both
quadratures of the response, minimizing their distance respectively from $\textrm{Re}X$ and $\textrm{Im}X$. The function obtained, expressed as amplitude and phase, is reported in Fig.~(\ref{risposta}) and the fitting parameters are summarized in Table~\ref{Tab1}. The factor $A$ is fixed to $32\cos^2\theta$; as we will show later (in Eq.~(\ref{LmodeFFP}) and the following discussion), this coefficient allows to attribute the modal masses $M_i$ to vibrations of the test masses, excited and read on the laser spots.
\begin{table}[t]
\caption{Parameters obtained by fitting the experimental response to a modulated intracavity power with expression (\ref{fitfun}).}
\label{Tab1}
\begin{center}
\begin{tabular}{lcc}
\hline \hline
$M_1$ (Kg) & \multicolumn{2}{c}{0.20} \\
$M_2$ (Kg) & \multicolumn{2}{c}{0.34} \\
$M_3$ (Kg) & \multicolumn{2}{c}{1.07} \\
$M_4$ (Kg) & \multicolumn{2}{c}{4.4} \\
$\nu_1$ (KHz) & \multicolumn{2}{c}{0.795} \\
$\nu_0$ (KHz) & \multicolumn{2}{c}{1.191} \\
$\nu_0$ (KHz) & \multicolumn{2}{c}{1.171} \\
$\nu_0$ (KHz) & \multicolumn{2}{c}{1.140} \\
$Q_1$ & \multicolumn{2}{c}{1940} \\
$Q_2$ & \multicolumn{2}{c}{835} \\
$Q_3$ & \multicolumn{2}{c}{370} \\
$Q_3$ & \multicolumn{2}{c}{70} \\
$\vert C_{phth} \vert$ (m/W) & \multicolumn{2}{c}{1.34~$10^{-13}$} \\
Phase($C_{phth}$) ($^0$) & \multicolumn{2}{c}{23} \\
\hline \hline
\end{tabular}
\end{center}
\end{table}

For the translation modes (at 795~Hz and 1190~Hz) we infer
effective masses of 0.2~kg and 0.34~kg respectively. While for the
latter the agreement with the physical mass and FEM simulation is
good, for the former it is only marginal. The effective mass of
the torsion mode at 1166~Hz is about 1~kg, much more than the
physical mass involved in the motion, showing that, as expected,
the readout is nearly blind to this mode. Also this effect is
particularly important for DUAL, where the readout topology is
studied in order to minimize the effect of disturbing mechanical
modes within the useful detection band~\cite{Bonaldi1}.

Concerning the mechanical quality factors, the values obtained
from the fit are respectively 1940 (at 795~Hz), 835 (at 1191 Hz),
and 370 (at 1171~Hz), all in good agreement with the corresponding
figures given by the displacement noise spectrum. The fourth
resonance (around 1140~Hz) is barely visible.

\begin{figure}[t]
\begin{center}
\includegraphics*[width=0.9\columnwidth]{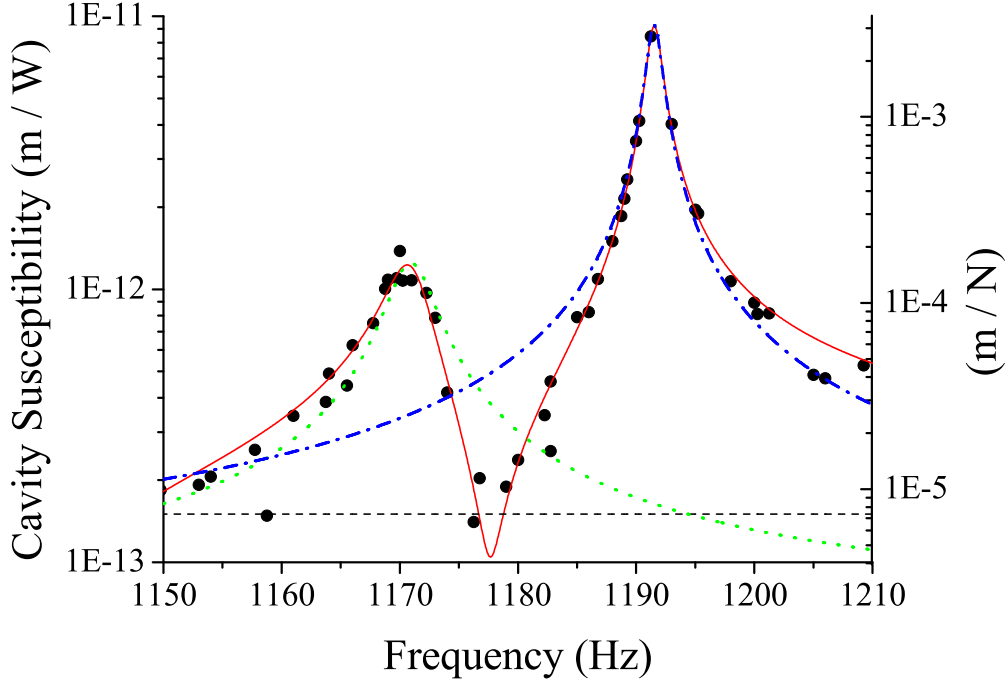}
\end{center}
\caption{Enlargement of the response shown in
Fig.~(\ref{risposta}), together with a complete fitting function
(red, solid line), the contributions of two peaks (respectively
green dotted and blue dashed-dotted lines) and contribution of the
photo-thermal background (gray dashed line).} \label{rispostazoom}
\end{figure}

It is particularly interesting the region around 1.2~kHz, shown in
detail in Fig.~(\ref{rispostazoom}), where a torsion and a
translation mode are very close. Here, between the two peaks, the
complete response falls below each single peak contribution (shown
with dashed lines in the figure), due to their interference. This
effect implies a cancellation of the back-action: the system is
less sensitive to modulation of the intracavity power, and
therefore also to laser amplitude fluctuations and radiation
shot-noise, which are the physical origin of the back-action. On
the other hand, the effect of a force acting on both modes in the
same direction would be amplified. The same should happen between the two gw sensitive modes of a DUAL detector. The demonstration of this effect on our prototype would be even clearer by exploiting the translation modes of the two masses, for which the reaction to a gw is more intuitive, but it is prevented by the photo-thermal background.

Let us now consider the other crucial issue of DUAL: the 'wide
area readout'. In the case of optical readout, it can be
implemented by the multi-spot configuration of the FFP. We first
summarize its working principle, in the simple scheme of $N$ spots
on one surface and $N+1$ on the opposite one. We suppose that the
spots are far enough to neglect the overlap between local
deformations, and the beam impinges with small incidence angle
(more detailed studies are reported in Ref.~\cite{Marin1}). In
this case, if the first surface moves by $\delta D$, the change in
the optical path is $\delta L = 2 N \delta D$, with an increase by
a factor of $2N$ with respect to a simple cavity. The radiation
pressure noise, due to intracavity power fluctuations, acts at the
same time on all the spots. The resulting displacement noise power
in the optical path is $S_{rp}^L = 4N^2 S_{rp}^x$, where
$S_{rp}^x$ are the fluctuations in the position of a single spot.
It seems that the signal-to-noise remains unchanged with respect
to a simple cavity, with both noise and signal power multiplied by
$4N^2$. However, we must consider that in a FFP the intracavity
power is reduced, and therefore also its fluctuations. Indeed, the
intracavity power is $P_c=\frac{T}{(T+A_{tot})^2}P_{in}$ where $T$
is the input mirror transmission, $A_{tot}$ the other roundtrip
losses, $P_{in}$ the input power. For an optimally coupled cavity
(the best configuration for a PDH detection), with $T = A_{tot}$,
one has $P_c=P_{in}/4A_{tot}$. Now, in a FFP, $A_{tot}= 2N
A_{single}$ where $A_{single}$ are the losses in a single spot (in
the round-trip, each spot is touched twice). Therefore the
intracavity power is reduced by a factor of $2N$ and its power
fluctuations by $4N^2$, as well as $S_{rp}^x$. As a consequence,
the signal-to-noise power ratio is improved by a factor $4N^2$.

The calculation of the Brownian thermal noise is even simpler: we
have uncorrelated noise fluctuations in the different spots,
therefore $S_{th}^L=4N S_{th}^x$ ($S_{th}^L$ are the fluctuations
in the FFP optical path, $S_{th}^x$ the displacement thermal noise
in a single spot) and the signal-to-noise is improved by a factor
of $N$.

In this work the sensitivity is not enough to see the effects of
'local' thermal noise and radiation pressure effects: the former
is overwhelmed by laser frequency noise, the latter by
photo-thermal background. However, for what the 'wide
area readout' is concerned, such photo-thermal expansion behaves exactly like a
deformation due to radiation pressure: phase and amplitude of the
response are different, but the effects on the different spots sum
up in the same way.

As a further difference with respect to the described model, we
cannot optimize the input mirror transmission. However, we can
measure  the ratio
between mechanical modes peaks and photo-thermal background in the response to modulated intracavity power.
Thanks to the broad area readout, the peaks emerge from
background stronger than in the case of a simple cavity. In
particular, for our configuration, if $\delta x = \chi_{phth}
\delta P_c$ is the photo-thermal displacement of each single spot
due to a change $\delta P_c$ in the intracavity power, the change
in the FFP optical length is
\begin{equation}
\label{dLphthFFP} \delta L^{FFP}_{phth} = 2(1+7\cos \theta)
\chi_{phth} \delta P_c   \, .
\end{equation}
This is illustrated in Fig.~(\ref{spostamenti}): on the end mirrors the change of optical length is $\delta x$, while on the seven folding mirrors it is $2\cos\theta\,\delta x$.
In the case of a simple cavity, summing the effects on its two
mirrors, we would have $\delta L^{simple}_{phth} = 2 \chi_{phth}
\delta P_c$. 

Due to a translation mode of one FFP mass, with
susceptibility at resonance $\chi_0$, we have instead
\begin{equation}
\label{LmodeFFP}
\delta L^{FFP}_{mode} = \frac{2}{c} (32\cos^2\theta) \chi_0
\,\delta P_c   \, ,
\end{equation}
to be compared with the simple cavity expression
\begin{equation}
\delta L^{simple}_{mode} = \frac{2}{c} \chi_0 \,\delta P_c   \, .
\end{equation}
Equation (\ref{LmodeFFP}) is obtained by using Eq.~(\ref{DtoL}) and the mass displacement $\delta D = \chi_0 \, \delta F$ where $\delta F = 4(\frac{2}{c}\cos\theta \, \delta P_c)$ due to the four bounces.
We finally derive
\begin{equation}
\frac{\delta L^{FFP}_{mode}}{\delta L^{FFP}_{phth}} =
\frac{2}{c}\,\frac{32\cos^2\theta}{2(1+7\cos\theta)}\,\frac{\chi_0}{\chi_{phth}}
\end{equation}
for the FFP, and
\begin{equation}
\frac{\delta L^{simple}_{mode}}{\delta L^{simple}_{phth}} =
\frac{2}{c}\,\frac{\chi_0}{2\chi_{phth}}
\end{equation}
for a simple cavity.

We have measured the photo-thermal expansion in a simple
Fabry-Perot cavity with the same mirrors used in the FFP and
roughly the same beam size (the cavity length is
0.2~m)\cite{DeRosa06}. In the frequency range of interest (around
1~kHz) the effect has a weak dependence on the frequency, and for
each single mirror is $\left|\chi_{phth}\right|=10^{-14}$~m/W. Using this value
in the expression~(\ref{dLphthFFP}), we infer  $\delta
L^{FFP}_{phth}/\delta P_c \simeq 1.5\cdot10^{-13}$~m/W, in good
agreement with the value of $1.3\cdot10^{-13}$~m/W obtained from
the fit of the experimental data (reported with a dashed line in
Fig.~(\ref{rispostazoom})).

Concerning the peak-to-background ratio, taking for instance the
first translation mode at 800~Hz with $\chi_0=3.9\cdot10^{-4}
m/N$, we calculate $\delta L^{FFP}_{mode}/\delta L^{FFP}_{phth} =
520$ and $\delta L^{simple}_{mode}/\delta L^{simple}_{phth} =
130$, to be compared with the experimental value of 540.

The data presented here can be considered as the first
experimental verification of the 'wide area readout' effect, for
the particular case of multi-spot readout.

\section{Radiation pressure cooling}

The experimental scheme for the radiation pressure cooling of a
FFP translation mode is similar to the setup implemented for the
measurement of the response to modulated intracavity power. The
measurement beam is weakly locked to a resonance of the optical
cavity, and its PDH signal is used also for obtaining the spectrum
of the cavity length fluctuations. The calibration of the spectral
measurement is performed with the procedure described in Section
(\ref{Expsetup}), even if now we are just interested in a narrow
frequency region around the peak at $\sim$800~Hz. The same PDH
signal, amplified and integrated, is used as correction signal
acting on the second beam for the active radiation pressure
cooling.

We have experimented two different cooling schemes. In the first
case, the second beam is tuned on resonance of its polarization
mode, and we act on its intensity (by means of EOM2). This
configuration implements a standard optical 'cold damping', like
the one firstly demonstrated in Ref.~\cite{Cohadon99} and later,
i.e., in Ref.~\cite{Kleckner06}. With respect to the mentioned
works, here we have a close spatial matching between sensing and
cooling beams.

In the second scheme, the cooling beam is frequency shifted on the
average by half linewidth from resonance, and the feedback acts on
the beam detuning (by means of one AOM) which actually controls the
intracavity power. This configuration is somehow original: its
working principle is the same as in Ref.~\cite{Mow08} (the
so-called hybrid scheme), with the difference that here we react
on the laser-to-cavity detuning by changing the laser frequency
instead of the cavity length. An important characteristics that distinguishes our experiment from the those described in Refs.~\cite{Mow08} and \cite{Corbitt} is the use of two separate laser beams for measuring the mass motion and for cooling. In this way, the measured signal (i.e., the detuning between the probe beam and the cavity) is not directly modified by the cooling servo-loop, that only acts on the frequency of the second beam without directly modifying the probe frequency, nor the cavity length. This procedure eliminate the necessity to correct the acquired data for the feedback: the probe beam can be considered as free-running and the PDH signal gives directly the cavity length fluctuations, thus allowing a clearer interpretation of the results.

The physics behind active cooling is described in
Refs.~\cite{Mancini}, \cite{Cohadon99} and in several subsequent
articles. We summarize here its main features, also in order to
clarify our experiment.

In general, a signal proportional to the detuning is frequency
filtered and transformed into a force acting on the oscillator.
The evolution equation of the oscillator position, in the Fourier
space, reads
\begin{equation}
\label{cool1} 
M\left(\omega_0^2-\omega^2-i
\frac{\omega\omega_0}{Q}\right)\tilde{x}=\tilde{f}_{th}+G\tilde{x}
\end{equation}
where G is a general complex, frequency-dependent gain function and $\tilde{f}_{th}$ is the noise force, that in the case of Brownian thermal noise has spectral density\cite{Saulson}
\begin{equation}
S_{th}=\frac{4 k T M \omega_0}{Q}  \, .
\end{equation}
A general, white force noise can be included by replacing the thermodynamic temperature $T$ with a noise temperature.
Eq.~(\ref{cool1}) gives
\begin{equation}
\tilde{x}=\frac{\tilde{f}_{th}}{M\left(\omega_{eff}^2-\omega^2- i
\frac{\omega\omega_{eff}}{Q_{eff}}\right)}
\end{equation}
with $\omega_{eff}^2=\omega_0^2-$Re$G/M$ and
$\frac{\omega_{eff}}{Q_{eff}}=\frac{\omega_0}{Q}+\frac{\textrm{Im}G}{M\omega}$.

Taking a feedback proportional to the oscillator velocity, e.g.,
$G=-i gM\omega\omega_0$, the effective eigenfrequency and quality
factor do not depend on $\omega$, and the equation of motion is
the same as the one of a free oscillator, with modified damping
($1/Q_{eff}=1/Q-g$) and temperature. In particular, for $g>0$ the
quality factor is increased and the effective temperature,
obtained from the integral of the spectrum as in Eq.~(\ref{Teff}),
is increased as $T_{eff}=TQ_{eff}/Q=T/(1-Qg)$. The same
expressions for negative $g$ gives lowering quality factor and
temperature ('cold damping').

If $G$ has a different dependence on $\omega$, the parallelism
between the modified evolution equation and a free oscillator is
not straightforward. Even in simple realistic schemes the energy
equipartition does not hold and a system temperature is not well
defined, as discussed, e.g., in Ref.~\cite{Genes}. However, for
high quality factor, most of the fluctuations are localized in a
narrow spectral region around resonance and we can use a
Lorentzian approximation of the spectrum, so that
\begin{equation}
S_x \simeq \frac{ S_{th} }{M^2 \,\omega_{eff}^2
\left[4(\omega_{eff}-\omega)^2+\left(\frac{\omega_{eff}}{Q_{eff}}\right)^2\right]}
\end{equation}
where in $\omega_{eff}$ and $Q_{eff}$ we can replace $G(\omega)$
by its value at $\omega_0$.  Even a colored external noise can be included in the discussion, provided that it has a smooth spectrum in the region of the peak. In this case an effective temperature
can still be defined according to Eq.~(\ref{Teff}), that gives
\begin{equation}
\label{Teff}
T_{eff}=\frac{T}{1+Q\frac{\textrm{Im}G(\omega_0)}{M \omega_0^2}}
\, ,
\end{equation}
while the effective eigenfrequency is shifted by the real part of
the gain.

In the adiabatic limit, for frequencies well below the cavity
linewidth, the force exerted by radiation pressure can be written
as
\begin{equation}
\label{Frp} F_{rp}=\frac{2}{c}\frac{P_c}{1+\Delta^2}
\end{equation}
where $P_c$ is the intracavity power at resonance and $\Delta$ the
detuning normalized to the cavity half-linewidth $\gamma$. 

In the
standard 'cold damping' scheme, the laser is kept at resonance and
a signal proportional to the detuning is sent to correct the laser
power. Neglecting the laser frequency fluctuations and the cavity
length noise (except for the oscillating mirror position $x$), the
loop gain can be written as
\begin{equation}
G=\frac{2}{c}P_c G_{el}
\end{equation}
where $G_{el}$ is a (complex and frequency-dependent) electronic
servo loop gain expressed as the ratio between the detuning
fluctuations $\tilde{x}$ and the consequent imposed relative power
fluctuations. 

In the hybrid configuration, the cooling laser is
detuned and the feedback is on the detuning. We can expand
Eq.~(\ref{Frp}) around the working point $\Delta_0$, obtaining for
the radiation pressure force fluctuations $\tilde{f}_{rp}$
\begin{equation}
\label{frp} \tilde{f}_{rp}=-\frac{2}{c}P_c\,\frac{2
\Delta_0}{(1+\Delta_0)^2}\,\frac{1}{\gamma}\left(\tilde{x}+\tilde{l}+\frac{l_{cav}}{\nu_L}\,\tilde{\nu}_L\right)
\end{equation}
where $l_{cav}$ is the average cavity length, $\tilde{l}$ its
fluctuations, $\nu_L$ is the cooling laser frequency with fluctuations
$\tilde{\nu}_L$, and we have neglected laser amplitude noise. The
cavity and/or the laser frequency fluctuations may contain a term
proportional to $\tilde{x}$ through electronic servo loop gains
that we call respectively $G_l$ and $(l_{cav}/\nu_L) G_{\nu}$. This term is obtained from a measurement of the detuning between the cavity and a probe beam.
Eq.~(\ref{frp}) can now be written in an interesting form as
\begin{equation}
\label{frp2}
\tilde{f}_{rp}=-K_{os}\left[(1+G_l+G_{\nu})\,\tilde{x}+\tilde{n} \right]
\end{equation}
where
\begin{equation}
\label{Kos} K_{os}=\frac{4 P_c\,\Delta_0}{\gamma
c\,(1+\Delta_0)^2}
\end{equation}
is the optical spring rigidity~\cite{Brag} and 
$\tilde{n}$ is a general extra-noise term. The meaning of $K_{os}$ is clear when replacing
Eq.~(\ref{frp2}) in the expression for $\omega_{eff}$: the effect
of radiation pressure is modifying the oscillator spring rigidity
$(M\omega_0^2)$ according to $M\omega_{eff}^2=\textrm{Re}K_{os}+
M\omega_0^2$. An oscillator damping is imposed by the reactive
component of the optical spring. In the passive 'self-cooling' it
is given by the delay in the optical field buildup inside the
cavity (that we have neglected in the adiabatic approximation). In
the active scheme, it is obtained by the feedback that, as we see
in Eq.~(\ref{frp2}), 'modifies' the optical spring rigidity.

Concerning the extra-noise, it includes a) the cavity length noise $\tilde{l}^n$ entering both directly in $\tilde{l}$ and from the measurement exploited for cooling (i.e., through the servo-loop); b) the cooling laser frequency noise $\tilde{\nu}_L^n$; c) the probe beam frequency noise $\tilde{\nu}_{probe}$. We have therefore
\begin{equation}
\tilde{n}=\tilde{l}^n(1+G_l+G_{\nu})+\frac{l_{cav}}{\nu_L}\left[\tilde{\nu}_L^n+(G_l+G_{\nu})\tilde{\nu}_{probe}\right]  \, .
\end{equation}
The right end side of Eq.~(\ref{cool1}) becomes $\tilde{f}_{th}- K_{os}(1+ G_l+G_{\nu})\tilde{x}- K_{os}\tilde{n}$ and we are in the conditions of the previous general discussion by replacing $- K_{os}(1+ G_l+G_{\nu})\rightarrow G$ and adding to the force noise spectrum $S_{th}$ the additional term $ K_{os}^2\,S_n$, where $S_n$ is the spectral density of $\tilde{n}$. 

In our case, the dominant term in $S_n$ comes from the probe beam frequency noise $S_{\nu}$ and the feedback on the cooling beam frequency is characterized by $|G_{\nu}|\gg 1$. Therefore, we can write $S_n \simeq (l_{cav}/\nu_L)^2 \,|G|^2 S_{\nu}$ and, expressing the gain as $G=|G|\exp(i\phi)$, the equation for the effective temperature given in Eq.~(\ref{Teff}) can be written in the useful form  
\begin{equation}
\label{Teff2}
\frac{T_{eff}}{T} = \frac{1+a\left(\frac{P_{in}}{P_0}\right)^2}{1+\frac{P_{in}}{P_0}}
\end{equation}
where $P_{in}$ is the input power of the cooling beam, the normalization constant $P_0$ is 
\begin{equation}
P_0=\frac{\gamma c (1+\Delta_0)^2}{4\Delta_0}\frac{M\omega_0^2}{Q}\frac{P_{in}}{P_c}
\end{equation}
and the extra-noise coefficient $a$ is
\begin{equation}
a=\frac{l_{cav}^2}{\nu_L^2}\frac{S_{\nu}}{S_x(0)}\frac{1}{\sin^2 \phi} \, ,
\end{equation}
where $S_x(0)=S_{th}Q^2/(M^2\omega_0^4)$ is the displacement noise at resonance measured before cooling. We remark that, for the optimal feedback phase $\phi=\pi/2$, the value of $a$ can be inferred directly from the displacement noise spectrum: $S_x(0)$ is the peak value, and $l_{cav}^2 S_{\nu}/\nu_L^2$ is the background. The minimum effective temperature $T_{min}$, achievable for $P_{in}/P_0=\sqrt{1+1/a}-1$, is $T_{min}/T=2a(\sqrt{1+1/a}-1)\simeq 2\sqrt{a}$. 

\begin{figure}[t]
\begin{center}
\includegraphics*[width=0.9\columnwidth]{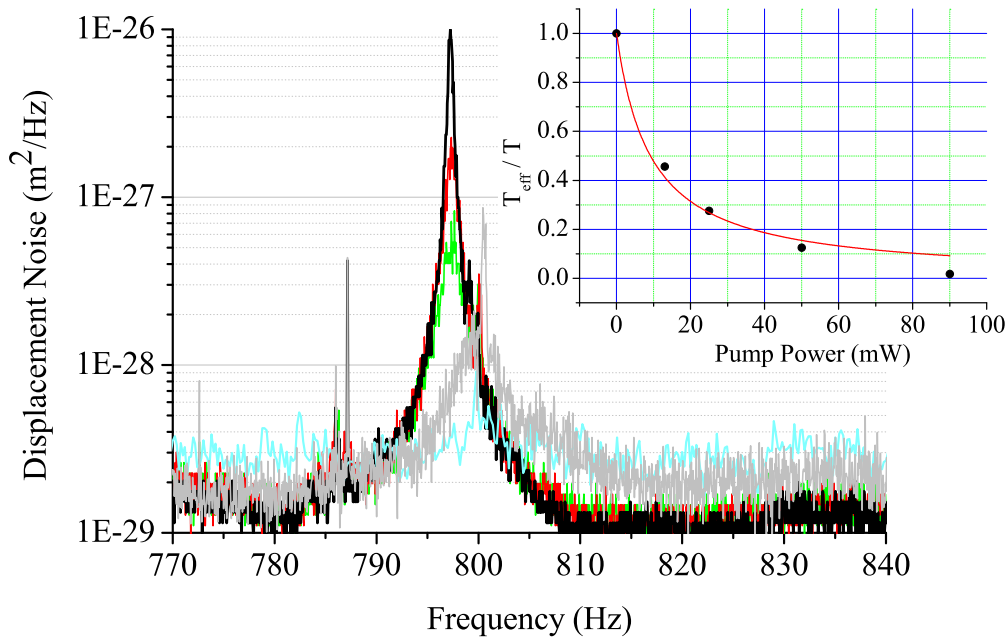}
\end{center}
\caption{Spectral measurement of the heavy mass displacement
noise, around the resonance of its lower translation mode, for
different values of the cooling laser power. From dark to light
lines, the corresponding laser power values measured before the
cavity are 0 (black line), 13~mW (red), 25~mW (green),
50~mW (light grey) and 90~mW (light blue). In the inset, the ratio between effective temperature $T_{eff}$ and initial temperature $T$ (symbols) as a function of the cooling laser power, fitted by Eq.~(\ref{Teff2}) (red line).} 
\label{cooling}
\end{figure}

We could achieve cooling of the translation mode both with 'cold damping' and with the hybrid scheme.
An example of the displacement noise spectra modified by cooling is reported in Fig.~(\ref{cooling}) for the second
configuration (with frequency feedback on the cooling beam). Here we focus on the translation mode of the heavy
mass, around 800~Hz, and we have transformed the cavity length
fluctuations into mass position fluctuations using the
relation~(\ref{DtoL}). The effective temperature of the mode,
estimated from the area of the spectral peak, is reduced by a
factor of more than 10, as shown in the inset of the figure. The cooling factor is limited the laser
frequency noise and, with strong cooling power, the
peak of the mechanical resonance disappears completely in the
frequency noise background. The measured effective temperature is fitted by Eq.~(\ref{Teff2}) fixing the extra-noise factor $a$ at the value of $1.1\cdot10^{-3}$, measured form the spectrum shown in Fig.~(\ref{spettro}). For the highest cooling power level, a weak instability in the electronic feedback loop slightly enhances the background noise. As a consequence, the uncertainty in the last temperature measurement is higher and we have not included in the fit, that gives $P_0=9.7$~mW and therefore a best obtainable cooling of $T_{min}/T=0.064$ for $P_{in}=280$~mW.

\section{Conclusions}

We have presented an experiment aimed at demonstrating and
testing the main new concepts on which are based the proposals of
DUAL gravitational wave detector. In particular, we show the
'back-action reduction' and the 'wide area readout' effects. Our
prototype also contains the first implementation of Folded
Fabry-Perot cavity, an optical scheme particularly conceived for
DUAL.

We report a complete characterization of the mechanical response
of $\sim$kg mass oscillators, performed by exciting the system
with radiation pressure force. Moreover, we describe active
optical cooling of an oscillator mode, also exploiting an
original scheme with feedback on the laser frequency. Such
experiences scale up the mass by several orders of magnitude with
respect to previous experiments on opto-mechanical effects, and focus on the
$\sim$kHz frequency region, of interest for large mass
gravitational wave detectors. Therefore, they represent a
significant step for realistic planning of the next generation of
massive detectors, whose possible optical readout should be pushed
to the quantum regime where radiation pressure effects are
critical\cite{Belfi}.

\section{Acknowledgments}

This work was partially funded by the European Union ILIAS Project (No. RII3-CT-2003-506222). We thank M. Cerdonio and M. De Rosa for useful discussions.


\begin{thebibliography}{99}
\bibitem{nota}We use in this article single-side spectra.
\end{thebibliography}
\end{document}